\documentstyle[12pt]{article}
\input math_macros.tex
\begin{document}
\begin{titlepage}
 \hfill   OHSTPY-HEP-T-94-010 (DOE/ER/01545-628) \\

  \vskip .1in

\center{\Large\bf SO(10) SUSY GUTs}
\center{\Large\bf and}
\center{\Large\bf Fermion Masses}

 \vskip .5in

\center{{Stuart Raby}\\{\it Department of Physics, The Ohio State
University}  \\{\it 174 W. 18th Ave., Columbus, OH 43210}\\{\it
raby@mps.ohio-state.edu}}

\vskip .5in

\flushleft{{\bf Abstract:}In this talk~\footnote{Talk presented at the
IFT Workshop on Yukawa Couplings, Gainesville, FL, February 1994.} I
summarize published work on a systematic operator analysis for fermion
masses in a class of effective supersymmetric SO(10)
GUTs\cite{adhrs}~\footnote{This work is in collaboration with G.
Anderson, S. Dimopoulos, L.J. Hall, and G. Starkman.}. Given a minimal
set of four operators at $M_G$, we have just 6 parameters in the
fermion mass matrices.  We thus make 8 predictions for the 14 low
energy observables (9 quark and charged lepton masses, 4 quark mixing
angles and $\tan \beta$).  Several models, i.e. particular sets of
dominant operators, are in quantitative agreement with the low energy
data.

 In the second half of the talk I discuss the necessary ingredients for
an SO(10) GUT valid below the Planck (or string) scale which reproduces
one of our models. \footnote{These are preliminary results of work in
progress with Lawrence Hall.}  This complete GUT should still be
interpreted as an effective field theory, i.e. perhaps the low energy
limit of a string theory.}

\end{titlepage}

\section{Introduction}

The Standard Model[SM] provides an excellent description of Nature.
Myriads of experimental tests have to date found no inconsistency.

Eighteen phenomenological parameters in the SM are necessary to fit all
the low energy data[LED].  These parameters are not equally well known.
$\alpha, \sin^2(\theta_W),$ $ m_e, m_{\mu}, m_{\tau}$ and $M_Z$ are all
known to better than 1\% accuracy.  On the otherhand,  $m_c, m_b,
V_{us}$ are known to between 1\% and 5\% accuracy,  and $
\alpha_s(M_Z), m_u, m_d,$ $ m_s, m _t,  V_{cb}, V_{ub}/V_{cb},
m_{Higgs}$ and the Jarlskog invariant measure of CP violation $J$ are
not known to better than 10\% accuracy.   One of the main goals of the
experimental high energy physics program in the next 5 to 10 years will
be to reduce these uncertainties.  In addition,  theoretical advances
in heavy quark physics and lattice gauge calculations will reduce the
theoretical uncertainties inherent in these parameters.  Already the
theoretical uncertainties in the determination of $V_{cb}$ from
inclusive B decays are thought to be as low as 5\%\cite{shifman}.
Moreover, lattice calculations are providing additional determinations
of $\alpha_s(M_Z)$ and heavy quark masses\cite{lattice}.

Accurate knowledge of these 18 parameters is important.  They are
clearly not a random set of numbers.  There are distinct patterns which
can, if we are fortunate, guide us towards a fundamental theory which
predicts some (if not all) of these parameters.  Conversely these 18
parameters are the LED which will test any such theory.  Note, that 13
of these parameters are in the fermion sector.  So, if we are to make
progress,  we must necessarily attack the problem of fermion masses.

In the program discussed in this talk, we define a procedure for
finding the dominant operator set reproducing the low energy data. In
the minimal operator sets we have just six parameters in the fermion
mass matrices.  We use the six best known low energy parameters as
input to fix these six unknowns and then predict the rest. These
theories are supersymmetric[SUSY] SO(10) grand unified theories[GUTs].
In the next two sections I want to briefly motivate these choices.

\section{Why SUSY GUTs?}

The answer has two parts.  First why SUSY?
\begin{enumerate}
\item  SUSY can provide at least a technical understanding to the
solution of the gauge hierarchy problem,  namely,  why  the ratio
$M_W/P_{Planck} \sim  10^{-16}$ is so small\cite{susy-gh}.
\item  It is also a very beautiful symmetry unifying states of
different spins\cite{susy}.  Clearly  many of the great successes of
particle physics in the last 30 years have involved the discovery of
one new symmetry of Nature after the next.  In fact all the forces of
the SM can be described in terms of local gauge symmetries (local SUSY
is supergravity\cite{sugra}).
\end{enumerate}

Now,  why GUTs?
\begin{enumerate}
\item  Once more this is a beautiful symmetry which unifies the known
forces of the SM and also the particles - quarks and leptons\cite{gg}.
The first consequence of this unification is that electric charge is
quantized in the correct way.  This success can also be claimed by
string theorists, without utilizing a GUT.
\item  More importantly, all GUTs make predictions.

\begin{enumerate}
\item  Given two gauge couplings measured at low energies,  the third
is predicted\cite{gqw}.  Thus using the experimentally measured values
of $\alpha$ and $sin^2\theta_W$,  the strong coupling $\alpha_s(M_Z)$
is fixed. This prediction is in good agreement with SUSY
GUTs\cite{drw1}.
\item  Nucleon decay is to be expected and in SUSY GUTs the dominant
decay modes of the proton are $p \rightarrow K^+ \bar{\nu}$ and $p
\rightarrow K^0 \mu^+$\cite{drw2}.
\end{enumerate}
In summary, only in SUSY GUTs are these predictions still consistent
with the low energy data.
\end{enumerate}

Note, as a direct consequence of a symmetry, in this case a SUSY GUT,
we have reduced the number of fundamental parameters in the gauge
sector of the theory by 1.  Consequently, we have gained one prediction
and this prediction agrees remarkably well with the LED\cite{drw1}.
This is a great success.  We also have the aesthetic feature of a SUSY
GUT DESERT, i.e. once the sparticles and gauginos are observed there is
no new physics until the GUT scale $M_G \sim 10^{16}$ GeV.  Thus there
is a huge discovery potential since now the LED can shed light directly
on the physics at $M_G$.

\section{Why SO(10)?}

There are two reasons for using SO(10).

\begin{enumerate}
\item  It is the smallest group in which all the fermions in one family
fit into one irreducible representation, i.e. the ${\bf 16}$.  Only one
additional state needs to be added to complete the multiplet and that
is a right-handed neutrino.  In larger gauge groups,  more as yet
unobserved states must be introduced to obtain complete multiplets.
Thus we take ${\bf 16}_i \supset \{ U_i, D_i, E_i, \nu_i \}, i = 1,2,3$
for the 3 families with the third family taken to be the heaviest.
{\em Since SO(10) Clebschs can now relate  $U, D, E$ and $\nu$ mass
matrices, we can in principle reduce the number of fundamental
parameters in the fermion sector of the theory.}  We return to this
point below.
\item  In any SUSY theory there are necessarily two higgs doublets --
$H_u$ and $H_d$.  Both these states fit into the ${\bf 10}$ of SO(10)
and thus their couplings to up and down type fermions are also given by
a Clebsch.  There are however six additional states in the ${\bf 10}$
which transform as a ${\bf 3}$ + ${\bf \overline{3}}$ under color.
These states contribute to proton decay and must thus be heavy.  The
problem of giving these color triplet states large mass of order $M_G$
while keeping the doublets light is sometimes called the second gauge
hierarchy problem.  This problem has a natural solution in SO(10) which
we discuss later\cite{dw}.
\end{enumerate}

Note that the gauge group SO(10) has to be spontaneously broken to the
gauge group of the SM -- $SU(3)\times SU(2)\times U(1)$.  This GUT
scale breaking can be accomplished by a set of states including \{
${\bf 45, 16, \overline{16}, \cdots }$\}.   The ${\bf 45}$(the adjoint
representation) enters into our construction of effective fermion mass
operators, thus I will discuss it in more detail in the next section.

I promised to return to the possibility of reducing the number of
fundamental parameters in the fermion sector of the theory.  Recall
that there are 13 such parameters.  Using symmetry arguments we can now
express the matrices ${\bf D, E},$ and ${\bf \nu}$ in terms of one
complex 3x3 matrix, $U$.  Unfortunately, this is not sufficient to
solve our problem.  There are 18 arbitrary parameters in this one
matrix.  In order to reduce the number of fundamental parameters we
must have zeros in this matrix.  We thus need new {\it family
symmetries} to enforce these zeros.

\section{The Big Picture}

Let us consider the big picture(see Fig. 1).  Our low energy observer
measures the physics at the electroweak scale and perhaps an order of
magnitude above.  Once the SUSY threshold is crossed we have direct
access to the effective theory at $M_G$, the scale where the 3 gauge
couplings meet.  Of course the GUT scale $M_G \sim 10^{16}$ GeV is
still one or two orders of magnitude below some more fundamental scale
such as the Planck or string scales (which we shall refer to as M).
Between M and $M_G$ there may be some substructure.  In fact, we may be
able to infer this substructure by studying fermion masses.

In our analysis we assume that the theory below the scale M is
described by a SUSY SO(10) GUT.  Between $M_G$ and $M$, at a scale
$v_{10}$, we assume that the gauge group SO(10) is broken spontaneously
to SU(5).  This can occur due to the vacuum expectation value of an
adjoint scalar in the X direction (i.e. corresponding to the U(1) which
commutes with SU(5)) and  the expectation values of a 16 and a
$\overline{16}$(denoted by $\Psi$ and $\overline{\Psi}$ respectively).
Then SU(5) is broken at the scale $v_5 = M_G$ to the SM gauge group.
This latter breaking can be done by different adjoints (45) in the Y,
B-L or T$_{3R}$ directions.

{\em Why consider 4 particular breaking directions for the 45 and no
others?}
The X and Y directions are orthogonal and span the two dimensional
space of U(1) subgroups of SO(10) which commute with the SM.  B-L and
T$_{3R}$ are also orthogonal and they span the same subspace.
Nevertheless we consider these four possible breaking directions and
these are the {\em only directions} which will enter the effective
operators for fermion masses.  Why not allow the X and Y directions or
any continous rotation of them in this 2d subspace of U(1) directions .
The answer is that there are good dynamical arguments for assuming that
these and only these directions are important. The X direction breaks
SO(10) to an intermediate SU(5) subgroup and it is reasonable to assume
that this occurs at a scale $v_{10} \ge v_5$.  Whether $v_{10}$ is
greater than $v_5$ or equal will be determined by the LED.  The B-L
direction is required for other reasons.  Recall the color triplet
higgs in the 10 which must necessarily receive large mass.  As shown by
Dimopoulos and Wilczek\cite{dw},  this doublet-triplet splitting can
naturally occur by introducing a 10 45 10 type coupling in the
superspace potential.  Note that the higgs triplets carry non-vanishing
B-L charge while the doublets carry zero charge.  Thus when the 45 gets
a vacuum expectation value[vev] in the B-L direction it will give mass
to the color triplet higgs at $v_5$ and leave the doublets massless.
Thus in any SO(10) model which solves this second hierarchy problem,
there must be a 45 pointing in the B-L direction.  We thus allow for
all 4 possible breaking vevs  --- X, Y, B-L and T$_{3R}$.  Furthermore
we believe this choice is ``natural" since we know how to construct
theories which have these directions as vacua without having to tune
any parameters.

Our fermion mass operators have dimension $\ge 4$.  {\em From where
would these higher dimension operators come?} Note that by measuring
the LED we directly probe the physics in some effective theory at
$M_G$.  This effective theory can, and likely will, include operators
with dimension greater than 4.     Consider, for example, our big
picture looking down from above.  String theories are very fundamental.
They can in principle describe physics at all scales.  Given a
particular string vacuum,  one can obtain an effective field theory
valid below the string scale M.  The massless sector can include the
gauge bosons of SO(10) with scalars in the 10, 45 or even 54
dimensional representations.  In addition, we require 3 families of
fermions in the 16.  Of course, in a string context when one says that
there are 3 families of fermions what is typically meant is that there
are 3 more 16s than $\overline{16}$s.  The extra 16 + $\overline{16}$
pairs are assumed to get mass at a scale $\ge M_G$,  since there is no
symmetry which prevents this.  When these states are integrated out in
order to define the effective field theory valid below $M_G$ they will
typically generate higher dimension operators.

Consider the tree diagram in Fig. 2.  The state  ${\bf 16_2}$ contains
the second generation of fermions.  It has off-diagonal couplings to
heavy fermions ${\bf 16_i, \overline{16}_i, i = 1,2}$.  If, for
example,  the vev ${\bf 45_X} > M_G$  then it will be responsible for
the dominant contribution to the mass of the states labelled 1 and 2.
These two states however mix by smaller off-diagonal mass terms given
by the vev ${\bf 45_{B-L}}$ or the singlet vev (  or mass term) denoted
by $M_G$.   When these heavy states are integrated out one generates to
leading order the operator  ${\bf O_{22}} =  16_2 10 {M_G 45_{B-L}
\over (45_X)^2} 16_2$  plus calculable corrections of order
$(v_5/v_{10})^2$.  {\em It is operators of this type (which can be
obtained by implicitly integrating out heavy 16's) which we use to
define our operator basis for fermion masses in the effective theory at
$M_G$.}

\section{Operator Basis for Fermion Masses at $M_G$}

Let us now consider the general {\bf operator basis for fermion
masses}.  We include operators of the form
$$
{\bf O_{ij}} = {\bf 16_i} ~( \cdots )_n ~{\bf 10} ~( \cdots )_m ~{\bf
16_j}  $$
where
$$( \cdots )_n =  {M_G^k ~45_{k+1} \cdots 45_n \over M_P^l
{}~45_X^{n-l}}
$$
and the $45$ vevs in the numerator can be in any of the 4 directions,
${\bf X, Y, B-L, T_{3R}}$ discussed earlier.

It is trivial to evaluate the
Clebsch-Gordon coefficients associated with any particular operator
since the matrices $X,Y,B-L,T_{3R}$ are diagonal.  Their eigenvalues on
the fermion states are given in Table 1.

\begin{table}[t]
\begin{center}
\begin{tabular}{|c|c|c|c|c|}
\multicolumn{5}{l}{Table~1. Quantum numbers of the}\\
 \multicolumn{5}{l}{four 45 vevs on fermion states.}  \\
\multicolumn{5}{l}{Note, if $u$ denotes a left-handed}\\
\multicolumn{5}{l}{up quark, then ${\bar u}$ denotes } \\
\multicolumn{5}{l}{a left-handed charge conjugate }\\
\multicolumn{5}{l}{up quark. }\\
\multicolumn{1}{c}{}&\multicolumn{4}{c}{}\\ \hline\hline
& ${\bf X}$ & ${\bf Y}$ & ${\bf B-L}$ & ${\bf T_{3R}}$
\\ \hline $u$ &  1 &  1/3 & 1 & 0  \\
${\bar u}$ &  1 & -4/3 & -1 & -1/2 \\
$d$ & 1 & 1/3 & 1 & 0 \\
${\bar d}$ & -3 & 2/3 & -1 & 1/2 \\
$e$ & -3 & -1 & -3 & 0 \\
${\bar e}$ & 1 & 2 & 3 & 1/2 \\
$\nu$ & -3 & -1 & -3 & 0 \\
${\bar \nu}$ & 5 & 0 & 3 & -1/2\\ \hline
\end{tabular}
\end{center}
\end{table}

\section{Dynamic Principles}

Now consider the dynamical principles which guide us towards a
theory of fermion masses.
\begin{description}
\item[0.] At zeroth order, we work in the context of a SUSY GUT with
the MSSM  below  $M_G$.
\item[1.] We use SO(10) as the GUT symmetry with three families of
fermions $\{ 16_i  ~~i = 1,2,3 \}$ and the minimal electroweak Higgs
content in one $10$.  SO(10) symmetry relations allow us to
reduce the number of fundamental parameters.
\item[2.] We assume that there are also family symmetries which enforce
zeros of the mass matrix,  although we will not specify these
symmetries at this time.  As we will make clear in section 12, these
symmetries will be realized at the level of the fundamental theory
defined below $M$.
\item[3.] Only the third generation obtains mass via
a dimension 4 operator.  The fermionic sector of the Lagrangian thus
contains the term $ A ~O_{33} \equiv A ~~16_3 ~10 {}~16_3$.  This term
gives mass to  t, b and $\tau$.  It results in the
symmetry relation --- $\lambda_t = \lambda_b = \lambda_{\tau} \equiv A$
at $M_G$.   This relation has been studied before by Ananthanarayan,
Lazarides and Shafi\cite{als} and using $m_b$ and $m_{\tau}$ as
input it leads to reasonable results for $m_t$ and $\tan \beta$.
\item[4.] All other masses come from operators with dimension $> 4$.
As a consequence,  the family hierarchy is related to the ratio
of scales above $M_G$.

\item[5.]  [{\bf Predictivity requirement}] ~We demand the
\underline{minimal set} of effective fermion mass operators at $M_G$
\underline{consistent with the {\bf LED}}.
\end{description}

\section{Systematic Search}

Our goal is to find the {\em minimal} set of fermion mass
operators consistent with the LED.  With any given operator set
one can evaluate the fermion mass matrices for up and down quarks and
charged leptons.  One obtains relations between mixing angles and
ratios of fermion masses which can be compared with the data.  It is
easy to show, however, without any detailed calculations that the
minimal operator set consistent with the LED is given by
\begin{eqnarray}
 & O_{33} + O_{23} + O_{22} + O_{12}&  ---
``22" ~{\rm texture} \nonumber\\
 {\rm or} & &  \nonumber\\
&  O_{33} + O_{23} + O'_{23} + O_{12}& --- ``23'" ~{\rm texture}
\nonumber
\end{eqnarray}

It is clear that at least 3 operators are needed to give
non-vanishing and unequal masses to all charged fermions, i.e. $ ~det
(m_a)  \neq 0$ for $a = u,d,e$.  That the operators must be in the [33,
23 and 12]
slots is not as obvious but is not difficult to show.  It is then
easy to show that 4 operators are required in order to have  CP
violation.  This is because, with only 3 SO(10) invariant operators,
we can redefine the phases of the three 16s of fermions to remove the
three arbitrary phases.  With one more operator, there is one
additional phase which cannot be removed.  A corollary of this
observation is that this minimal operator set results in just 5
arbitrary parameters in the Yukawa matrices of all fermions,  4
magnitudes and one phase\footnote{This is two fewer parameters than was
necessary in our previous analysis (see \cite{dhr})}.  This is the
minimal parameter set which can be obtained without solving the
remaining problems of the fermion mass hierarchy, one overall real
mixing angle and a CP violating
phase.  We should point out however that the problem of understanding
the fermion mass hierarchy and mixing has been rephrased as the problem
of understanding the hierarchy of scales above $M_G$.

{}From now on I will just consider models with ``22" texture.  This is
because they can reproduce the observed hierarchy of fermion masses
without fine-tuning\footnote{For more details on this point, see
section 9 below or refer to \cite{adhrs}.}.  Models with ``22" texture
give the following Yukawa matrices at $M_G$ (with electroweak doublet
fields on the right) --

$$
{\bf \lambda_a} = \left( \begin{array}{ccc}
0 & z'_a ~C & 0\\
z_a ~C & y_a ~E ~e^{i \phi} & x'_a ~B\\
0 & x_a ~B & A
\end{array} \right) $$
 with the subscript $a = \{ u, d, e\}$.  The constants
$x_a, x'_a, y_a, z_a, z'_a$ are Clebschs which can be determined once
the 3 operators ( $O_{23},O_{22}, O_{12}$) are specified.  Recall, we
have taken $O_{33} = A ~16_3 ~10 ~16_3$, which is why the Clebsch in
the 33 term is independent of $a$. Finally, combining the Yukawa
matrices with the Higgs vevs to find the fermion mass matrices we have
6 arbitrary parameters given by $A, B, C, E, \phi$ and $\tan \beta$
describing 14 observables.  We thus obtain 8 predictions.  We shall
use the best known parameters, $m_e, m_{\mu}, m_{\tau}, m_c, m_b,
|V_{cd}|$, as
input to fix the 6 unknowns.  We then predict the values of $m_u, m_d,
m_s,
m_t, \tan \beta, |V_{cb}|, |V_{ub}|$ and $J$.

Note: since the predictions are correlated, our analysis would be much
improved if we minimized some $\chi^2$ distribution and obtained a best
fit to the data.  Unfortunately this has not yet been done.  In the
paper however we do include some tables (see for example Table 4 in
this talk) which give all the predictions for a particular set of input
parameters.

\section{Results}

The results for the 3rd generation are given in Fig. 3.  Note that
since the parameter A is much bigger than the others we can essentially
treat the 3rd generation independently.  The small corrections, of
order $(B/A)^2$, are however included in the complete analysis. We find
the pole mass for the top quark $M_t = 180 \pm 15$ GeV and $\tan \beta
= 56\pm 6$ where the uncertainties result from variations of our input
values of the $\overline{MS}$ running mass $m_b(m_b) = 4.25 \pm 0.15$
and $\alpha_s(M_Z)$ taking values $.110 - .126$. We used two loop RG
equations for the MSSM from $M_G$ to $M_{SUSY}$;  introduced a
universal SUSY threshold at $M_{SUSY} = 180$ GeV with 3 loop QCD and 2
loop QED RG equations below $M_{SUSY}$.  The variation in the value of
$\alpha_s$ was included to indicate the sensitivity of our results to
threshold corrections which are necessarily present at the weak and GUT
scales.  In particular, we chose to vary $\alpha_s(M_Z)$ by letting
$\alpha_3(M_G)$ take on slightly different values than $\alpha_1(M_G) =
\alpha_2(M_G) = \alpha_G$.

The following set of operators passed a straightforward but coarse
grained search discussed in detail in the paper\cite{adhrs}.  They
include
the diagonal dimension four coupling of the third generation --
$$
\eqalignno{
    O_{33} = &  16_3\ 10\ 16_3 & \cr}
$$.

The six possible $O_{22}$ operators --
$$
\eqalignno{
O_{22}  =  &  &\cr
& 16_2 \ {45_X \over M} \ 10 \ {45_{B-L} \over 45_X} \ 16_2 & (a)\cr
& 16_2 \ {M_G \over \ 45_X} \ 10 \ {45_{B-L}\over M} \ 16_2 &(b)\cr
& 16_2 \ {45_X \over M} \ 10 \ {45_{B-L} \over M} \ 16_2 & (c)\cr
& 16_2 \ 10 \ {45_{B-L}\over 45_X} \  16_2 &(d)\cr
& 16_2 \ 10 \ {45_X \ 45_{B-L} \over M^2} \ 16_2  & (e)\cr
& 16_2 \ 10 \ {45_{B-L} \ M_G \over 45_X^2}\ 16_2 & (f)\cr}
$$
Note: in all cases the Clebschs $y_i$ ( defined by $O_{22}$ above)
satisfy

$$
y_u : y_d : y_e = 0 : 1 : 3.
$$
This is the form familiar from the Georgi-Jarlskog texture\cite{gj}.
Thus all six of these operators lead to {\it identical} low energy
predictions.

Finally there is a unique operator $O_{12}$ consistent with the LED --
$$
O_{12} = 16_1
\left( {45_X\over M}\right)^3 \ 10 \left( {45_X \over M}\right)^3
16_2
$$

The operator $O_{23}$  determines the KM element $V_{cb}$ by the
relation --
$$
V_{cb} = \chi \ \sqrt{ {m_c \over m_t}} \times ( RG factors)
$$
where the Clebsch combination $\chi$ is given by
$$
\chi \equiv {|x_u-x_d|\over \sqrt{|x_ux_u'|}}
$$
$m_c$ is input, $m_t$ has already been determined and the
renormalization group[RG] factors are calculable.  Demanding the
experimental constraint

$V_{cb} < .054$ we find the constraint $\chi < 1$.
A search of all operators of dimension 5 and 6 results in the 9
operators given below. Note that there only three different values of
$\chi = 2/3, ~5/6,~8/9$ --

$$
\eqalignno{ O_{23} = &  & \cr
 & & \chi = 2/3 \cr
(1)& 16_2 \ {45_{Y} \over M} \ 10 {M_G \over 45_X} \ 16_3 & \cr
(2)& 16_2 \ {45_{Y} \over M} \ 10\ {45_{B-L}\over 45_X} \ 16_3 & \cr
(3)& 16_2 \ {45_{Y}\over 45_X} \ 10 \ {M_G \over 45_X} \ 16_3 & \cr
(4)& 16_2 \ {45_{Y}\over 45_X} \ 10 \ {45_{B-L}\over45_X}   16_3 & \cr}
$$
$$
\eqalignno{
& & \chi = 5/6 \cr
(5)&  16_2 \ {45_{Y} \over M} \ 10 \ {45_{Y}\over 45_X} \ 16_3 & \cr
(6)&  16_2 \ {45_{Y}\over 45_X} \ 10 \ {45_{Y}\over 45_X} \ 16_3 & \cr}
$$
$$
\eqalignno{
& & \chi= 8/9 \cr
(7)&  16_2 \ 10 \ {M_G^2 \over 45_X^2} \ 16_3 & \cr
(8)&  16_2 \ 10 \ {45_{B-L} M_G \over 45_X^2} \ 16_3 & \cr
(9)&  16_2 \ 10 \ {45_{B-L}^2\over 45^2_X} \ 16_3  & \cr}
$$
We label the operators (1) - (9), and we use these numbers also to
denote the corresponding models.  {\em Note, all the operators have the
vev $45_X$ in the denominator.  This can only occur if  $v_{10} > M_G$.
}

At this point, there are no more simple criteria to reduce the number
of models further.  We have thus performed a numerical RG analysis on
each of the 9 models (represented by the 9 distinct operators $O_{23}$
with their calculable Clebschs $x_a, x'_a, a= u,d,e$  along with the
unique set of Clebschs determined by the operators $O_{33}, O_{22}$ and
$O_{12}$). We then iteratively fit the 6 arbitrary parameters to the
six low energy inputs and evaluate the predictions for each model as a
function of the input parameters.  The results of this analysis are
given in Figs. (4 - 10).

Let me make a few comments.  Light quark masses (u,d,s) are
$\overline{MS}$ masses evaluated at 1 GeV while heavy quark masses
(c,b) are evaluated at ($m_c, m_b$) respectively.  Finally, the top
quark mass in Fig. 3 is the pole mass.  Figs. 4 and 5 are self evident.
In Fig. 6, we show the correlations for two of our predictions.  The
ellipse in the $m_s/m_d$ vs. $m_u/m_d$ plane is the allowed region from
chiral Lagrangian analysis\cite{chiral}. One sees that we favor lower
values of $\alpha_s(M_Z)$. For each fixed value of $\alpha_s(M_Z)$,
there are 5 vertical line segments in the $V_{cb}$ vs. $m_u/m_d$ plane.
Each vertical line segment represents a range of values for $m_c$ (with
$m_c$ increasing moving up) and the different line segments represent
different values of $m_b$ (with $m_b$ increasing moving to the left).
In Figure 9 we test our agreement with the observed CP violation in the
K system.  The experimentally determined value of $\epsilon_K = 2.26
\times 10^{-3}$. Theoretically it is given by an expression of the form
$B_K \times \{ m_t, V_{ts}, \cdots \}$.   $B_K$ is the so-called Bag
constant which has been determined by lattice calculations to be in the
range $B_K = .7 \pm .2$\cite{bag}.  In Fig. 9 we have used our
predictions for fermion masses and mixing angles as input, along with
the experimental value for $\epsilon_K$, and fixed $B_K$ for the 9
different models.  One sees that model 4 is inconsistent with the
lattice data.  In Fig. 10 we present the predictions for each model,
for the CP violating angles which can be measured in B decays.  The
interior of the ``whale" is the range of parameters consistent with the
SM found by Nir and Sarid\cite{ns} and the error bars represent the
accuracy expected from a B factory.

Note that model 4 appears to give too little CP violation and model 9
has uncomfortably large values of $V_{cb}$.  Thus these models are
presently disfavored by the data.  I will thus focus on model 6 in the
second part of this talk.

\section{Summary}

We have performed a systematic operator analysis of fermion masses in
an effective SUSY SO(10) GUT.  We use the LED to lead us to the theory.
Presently there are 3 models (models 4, 6 \& 9) with ``22" texture
which agree best with the LED, although as mentioned above model 6 is
favored.  In all cases we used the values of $\alpha$ and
$\sin^2\theta_W$ (modulo threshold corrections) to fix
$\alpha_s(M_Z)$.

Table 2 shows the virtue of the ``22" texture.  In the first column are
the four operators.  In the 2nd and 3rd columns are the parameters in
the mass matrix relevant for that particular operator and the input
parameters which are used to fix these parameters.  Finally the 4th
column contains the predictions obtained at each level.  One sees that
each family is most sensitive to a different operator\footnote{This
property is not true of ``23" textures.}.

\begin{table}[t]
\begin{center}
\begin{tabular}{|c|c|c|c|}
\multicolumn{4}{l}{Table~2. Virtue of ``22" texture.}\\
 \multicolumn{4}{c}{}\\ \hline\hline
Operator  & Parameters & Input & Predictions \\ \hline

$O_{33}$ &  $\tan\beta$ \ A &  b \ $\tau$ & t  \ $\tan\beta$  \\
$O_{23}$ &  B & c & $V_{cb}$  \\
$O_{22}$ & E & $\mu$ & s  \\
$O_{12}$ & C \ $\phi$  & e \ $V_{us}$ & u \ d \ ${V_{ub} \over V_{cb}}$
\ J \\ \hline
\end{tabular}
\end{center}
\end{table}

{\em Consider the theoretical uncertainties inherent in our analysis.}
\begin{enumerate}
\item The experimentally determined values of $m_b, m_c$, and
$\alpha_s(M_Z)$ are all subject to strong interaction uncertainties of
QCD.  In addition, the predicted value of $\alpha_s(M_Z)$ from GUTs is
subject to threshold corrections at $M_W$  which can only be calculated
once the SUSY spectrum is known and at $M_G$ which requires knowledge
of the theory above $M_G$.  We have included these uncertainties
(albeit crudely) explicitly in our analysis.

\item In the large $\tan \beta$ regime in which we work there may be
large SUSY loop corrections which will affect our results.  The finite
corrections to the $b$ and $\tau$ Yukawa couplings have been
evaluated\cite{hrs,copw}.  They depend on ratios of soft SUSY breaking
parameters and are significant in certain regions of parameter
space\footnote{There is a small range of parameter space in which our
results are unchanged\cite{hrs}.  This requires threshold corrections
at $M_G$ which distinguish the two Higgs scalars.}.  In particular it
has been shown that the top quark mass can be reduced by as much as
30\%.  Note that although the prediction of Fig. 3 may no longer be
valid,  there is still necessarily a prediction for the top quark mass.
It is now however sensitive to the details of the sparticle spectrum
and to the process of radiative electroweak symmetry
breaking\cite{resb}.  This means that the observed top quark mass can
now be used to set limits on the sparticle spectrum.   This analysis
has not been done.  Moreover,  there are also similar corrections to
the Yukawa couplings for the $s$ and  $d$ quarks and for $e$ and $\mu$.
These corrections are expected to affect the predictions for $V_{cb},
m_s, m_u, m_d$.   It will be interesting to see the results of this
analysis.

\item The top, bottom and $\tau$ Yukawa couplings can receive threshold
corrections at $M_G$.  We have not studied the sensitivity to these
corrections.

\item Other operators could in principle be added to our effective
theory at $M_G$.  They might have a dynamical origin. We have assumed
that, if there, they are subdominant.  Two different origins for these
operators can be imagined.  The first is field theoretic. The operators
we use would only be the leading terms in a power series expansion when
defining an effective theory at $M_G$ by integrating out heavier
states.  The corrections to these operators are expected to be about
10\%.   We may also be sensitive to what has commonly been referred to
as Planck slop\cite{pslop}, operators suppressed by some power of the
Planck (or string) scale M.   In fact the operator $O_{12}$ may be
thought of as such.  The question is why aren't our results for the
first and perhaps the second generation,  hopelessly sensitive to this
unknown physics?  This question will be addressed in the next section.
\end{enumerate}

\section{Where are we going?}

In the first half of Table 3 I give a brief summary of the good and bad
features of the effective SUSY GUT  discussed earlier.  Several models
were found with just four operators at $M_G$ which successfully fit the
low energy data. If we add up all the necessary parameters needed in
these models we find just 12.  This should be compared to the SM with
18 or the MSSM with 21.  Thus these theories,  minimal effective SUSY
GUTs[MESG], are doing quite well.  Of course the bad features of the
MESG is that it is not a fundamental theory.  In particular there are
no symmetries which prevent additional higher dimension operators to
spoil our results.  Neither are we able to calculate threshold
corrections, even in principle, at $M_G$.

It is for these reasons that we need to be able to take the MESG which
best describes the LED and use it to define an effective field theory
valid at scales $\le M$.  The good and bad features of the resulting
theory are listed in the second half of Table 3.

\begin{itemize}
\item In the effective field theory below $M$ we must incorporate the
{\it symmetries} which guarantee that we reproduce the MESG with no
additional operators\footnote{This statement excludes the unavoidable
higher order field theoretic corrections to the MESG which are, in
principle, calculable.}

{\em Moreover, the necessary combination of discrete, U(1) or R
symmetries may be powerful enough to restrict the appearance of Planck
slop.}

\item Finally,  the {\it GUT symmetry breaking} sector must resolve the
problems of natural doublet-triplet splitting (the second hierarchy
problem), the $\mu$ problem, and give predictions for proton decay,
neutrino masses and calculable threshold corrections at $M_G$.

\item On the bad side,  it is still not a fundamental theory and there
may not be a unique extension of the MESG to higher energies.
\end{itemize}

\begin{table}[t]
\begin{center}
\begin{tabular}{|c|c|c|}
\multicolumn{3}{l}{Table~3. }\\
 \multicolumn{3}{l}{}\\ \hline\hline
   & Good & Bad \\ \hline

   Eff. F.T.  & \underline{ 4 op's. at $M_G \Rightarrow$ {\bf LED}}&
\underline{Not fundamental} \\
  $\le M_G$ & 5 para's. $\Rightarrow$ 13 observables & \underline{No
symmetry}  \\
  & + 2 gauge para's. $\Rightarrow$ 3 observables & $\Rightarrow$ Why
these operators?  \\
    & \underline{+ 5} soft SUSY  &  (F.T. + Planck slop) \\

    &  breaking para's. $\Rightarrow \cdots$ &      \\

  &  Total {\bf 12} parameters &  \underline{Threshold corrections?} \\
\hline
Eff. F.T.  &  \underline{Symmetry} &  \underline{Not fundamental} \\
  $\le M $ & i)  gives Eff. F.T. $\le M_G$  &    \\
$M = M_{string}$ &         + corrections          &   \underline{Not
unique?} \\
 or $M_{Planck}$ & ii)  constrains other operators  &          \\
   &   \underline{GUT symmetry breaking}  &      \\
  &  i)  d - t splitting   &         \\
   &  ii)  $\mu$ problem   &          \\
    &  iii)  proton decay   &        \\
    &  iv)  neutrino masses  &        \\
     &  v)  threshold corrections at $M_G$  &     \\  \hline
\end{tabular}
\end{center}
\end{table}

\section{String Threshold at $M_S$}

Upon constructing the effective field theory $\le M_S$, we will have
determined the necessary SO(10) states, symmetries and couplings which
reproduce our fermion mass relations.  This theory can be the starting
point for constructing a realistic string model.  String model builders
could try to obtain a string vacuum with a massless spectra which
agrees with ours.  Of course, once the states are found the string will
determine the symmetries and couplings of the theory.   It is hoped
that in this way a {\it fundamental} theory of Nature can be found.
Recent work in this direction has been reported in this workshop by
Joseph Lykken\cite{lykken}.  He has been able to obtain string theories
with SO(10), three families plus additional 16 + $\overline{16}$ pairs,
45's, 10's and even some 54 dimensional representations.  One of the
first results from this approach is the fact that only one of the three
families has diagonal couplings to the 10,  just as we have assumed.

\section{Constructing the Effective Field Theory below $M_S$}

In this section I will discuss some preliminary results obtained in
collaboration with Lawrence Hall. I will describe the necessary
ingredients for constructing model 6.  Some very general results from
this exercise are already apparent.

\begin{itemize}
\item {\it States}  We have constructed a SUSY GUT which includes all
the states necessary for GUT symmetry breaking and also for generating
the 45 vevs in the desired directions.  A minimal representation
content below $M_S$ includes 54s + 45s + 3  16s + n($\overline{16}$ +
16) pairs  + 2  10s.
\item {\it Symmetry}  In order to retain sufficient symmetry the
superspace potential in the visible sector W necessarily has a number
of flat directions.  In particular the scales $v_5$ and $v_{10}$ can
only be determined when soft SUSY breaking and quantum corrections are
included.  An auxiliary consequence is that the vev of W$_{visible}$
vanishes in the supersymmetric limit.
\item {\it Couplings}  As an example of the new physics which results
from this analysis I will show how a solution to the $\mu$ problem, the
ratio $\lambda_b/\lambda_t$ and proton decay are inter-related.
\end{itemize}

In Table 4 are presented the predictions for Model 6 for particular
values of the input parameters.

\begin{center}
\indent Table 4: Particular Predictions for Model 6
with $\alpha_s(M_Z) = 0.115$
\vskip 20pt
\begin{tabular}{|c|c|c|c|}
\hline
  Input  & Input & Predicted & Predicted  \cr
 Quantity &  Value &  Quantity & Value \cr
\hline
  $m_b(m_b)$ & $4.35 $ GeV & $M_t$ & $176$ GeV \cr
  $m_\tau(m_\tau)$ &$1.777 $ GeV & $\tan\beta$ & $55 $  \cr
\hline
  $m_c(m_c)$ &$1.22$ GeV & $V_{cb}$ & $.048 $ \cr
\hline
  $m_\mu$ &$105.6 $ MeV   & $V_{ub}/V_{cb}$ & $.059 $ \cr
  $m_e $   &$0.511$ MeV  &  $m_s(1GeV)$ & $172 $ MeV \cr
  $V_{us}$       &$0.221 $    & $\hat{B_K} $ & $0.64$  \cr
                 &            & $m_u / m_d$  & $0.64$  \cr
                 &            & $m_s / m_d$  & $24.$  \cr
\hline
\end{tabular}
\end{center}
In addition to these predictions, the set of inputs in
Table 4 predicts:

$\sin 2\alpha = -.46$, $\sin 2\beta = .49$,
$\sin 2\gamma = .84$, and $J = 2.6\times 10^{-5}$.

\begin{center}
{\Large\bf Model 6}
\end{center}

\vskip .1in

The superspace potential for Model 6 has several pieces -  W =
W$_{fermion}$ + W$_{symmetry \ breaking}$ + W$_{Higgs}$ +
W$_{neutrino}$.

\subsection{Fermion sector}

The first term must reproduce the four fermion mass operators of Model
6.  They are given by
$$
\eqalignno{
    O_{33} = &  16_3\ 10_1 \ 16_3 & \cr
     O_{23} = & 16_2 \ {A_2 \over {\tilde A}} \ 10_1 \ {A_2 \over
{\tilde A}} \ 16_3 & \cr
    O_{22}  =  &  16_2 \ {{\tilde A} \over M} \ 10_1 \ {A_1 \over
{\tilde A}} \ 16_2 & \cr
    O_{12} = & 16_1 \left( {{\tilde A}\over M}\right)^3 \ 10_1 \left(
{{\tilde A} \over M} \right)^3 16_2 & \cr}
$$

There are two 10s in this model, denoted by $10_i, i= 1,2$ and only
$10_1$ couples to the ordinary fermions.   The A fields are different
45s which are assumed to have vevs in the following directions --
$\vev{A_2} = 45_Y$,  $\vev{A_1} = 45_{B-L}$, and  $\vev{\tilde A} =
45_X$.  As noted earlier, there are 6 choices for the 22 operator and
we have just chosen one of them, labelled a, arbitrarily here.   In
Figure 11,  we give the tree diagrams which reproduce the effective
operators for Model 6 to leading order in an expansion in the ratio of
small to large scales.  The states  $\overline{\Psi}_a, \Psi_a, a = 1,
\cdots ,9$ are massive $\overline{16}, 16$ states respectively.   Each
vertex represents a separate Yukawa interaction in W$_{fermion}$ (see
below).  Field theoretic corrections to the effective GUT operators may
be obtained by diagonalizing the mass matrices for the heavy states and
integrating them out of the theory.

\begin{center}
\underline{\bf $W_{fermion}  $}
\end{center}

$$ 16_3 16_3 10_1  +  {\bar \Psi}_1 A_2 16_3  +  {\bar \Psi}_1 {\tilde
A} \Psi_1  +  \Psi_1 \Psi_2 10_1  $$

$$ +  {\bar \Psi}_2 {\tilde A} \Psi_2  +  {\bar \Psi}_2 A_2 16_2  +
{\bar \Psi}_3  A_1 16_2  $$
$$ +  {\bar \Psi}_3 {\tilde A} \Psi_3  +  \Psi_3 \Psi_4 10_1  +  M
\sum_{a=4}^9  ( {\bar \Psi}_a \Psi_a )  $$

$$ + {\bar \Psi}_4 {\tilde A} 16_2  +  {\bar \Psi}_5 {\tilde A} \Psi_4
+
{\bar \Psi}_6 {\tilde A} \Psi_5  $$

$$  +  \Psi_6 \Psi_7 10_1  +  {\bar \Psi}_7 {\tilde A} \Psi_8  +  {\bar
\Psi}_8 {\tilde A} \Psi_9  +  {\bar \Psi}_9 {\tilde A} 16_1 $$

\begin{itemize}
\item {\em Note that the vacuum insertions in the effective operators
above cannot be rearranged,  otherwise an inequivalent low energy
theory would result.  In order to preserve this order naturally we
demand that each field carries a different value of a U(1) family
charge (see Fig. 11).} Note also that the particular choice of a 22
operator will affect the allowed U(1) charges of the states.  Some
choices may be acceptable and others not.
\item Consider W$_{fermion}$.  It has many terms,  each of which can
have different, in principle, complex Yukawa couplings.  {\em
Nevertheless the theory is predictive because only a very special
linear combination of these parameters enters into the effective theory
at $M_G$.  Thus the observable low energy world is simple, not because
the full theory is particularly simple, but because the symmetries are
such that the effective low energy theory contains only a few dominant
terms.}
\end{itemize}

\subsection{Symmetry breaking sector}

The symmetry breaking sector of the theory is not particularly
illuminating.  Two 54 dimensional representations, $S, S'$ are needed
plus several singlets denoted by ${\cal S}_i, i = 1, \cdots, 7$.  They
appear in the first two terms and are responsible for driving the vev
of $A_1$ into the B-L direction,  the third term drives the vev of the
$\overline{16},16$ fields $\overline{\Psi}, \Psi$ into the right-handed
neutrino like direction breaking SO(10) to SU(5) and forcing ${\tilde
A}$ into the X direction.  The fourth and fifth term drives $A_2$ into
the Y direction. Finally the last two terms are necessary in order to
assure that all non singlet states under the SM gauge interactions
obtain mass of order the GUT scale.  All  primed fields are assumed to
have vanishing vevs.

Note if $\vev {\cal S}_3 \approx M_S$ then two of these adjoints state
may be heavy.  Considerations such as this will affect how couplings
run above $M_G$.

\begin{center}
\underline{\bf $W_{symmetry \,\, breaking}$}
\end{center}

$$    A'_1 (S A_1 + {\cal S}_1 A_1 ) +  S' ( {\cal S}_2 S + A_1 A_2) $$

$$  + {\tilde A}' ( {\bar \Psi} \Psi + {\cal S}_3  \tilde A ) $$

$$ +  A'_2 ( {\cal S}_4 A_2  +  S  {\tilde A} + ( {\cal S}_1 + {\cal
S}_5) {\tilde A} ) $$

$$ +   {\bar \Psi}' A_2 \Psi  +  {\bar \Psi} A_2 \Psi' $$

$$ +  A'_1 A_1 {\tilde A}  +  {\cal S}_6  {\tilde A}'^2  $$

\subsection{Higgs sector}

The Higgs sector is introduced below.  It does not at the moment appear
to be  unique, but it is crucial for understanding the solution to
several important problems -- doublet-triplet splitting,  $\mu$ problem
and proton decay -- and these constraints may only have one solution.
The $10_1 A_1 10_2$ coupling is  the term required by the
Dimopoulos-Wilczek mechanism for doublet-triplet splitting.  Since
$A_1$ is an anti-symmetric tensor, we need at least two 10s.

The couplings of $10_1$ to the 16s are introduced to solve the $\mu$
problem.  After naturally solving the doublet-triplet splitting problem
one has massless doublets.  One needs however a small supersymmetric
mass $\mu$ for the Higgs doublets of order the weak scale.  This may be
induced once SUSY is broken in several ways.

\begin{itemize}
\item The vev of the field $A_1$ may shift by an amount of order the
weak scale due to the introduction of the soft SUSY breaking terms into
the potential. In this theory the shift of $A_1$ appears to be too
small.
\item There may be higher dimension D terms in the theory of the form,
eg.
$${1 \over M_{Pl}}\int d^4\theta 10_1^2 (A_2^*)  $$.  Then supergravity
effects might induce a non-vanishing vev for the F term of $A_2$ of
order the $m_W M_G$.  This will induce a value of $\mu$ of order  $m_W
M_G/M_{Pl}$.  The shift in the F-terms also appear to be negligible.
\item One loop effects may induce a $\mu$ term once soft SUSY breaking
terms are introduced\cite{hall}.  In this case we find  $\mu \sim  {A
\lambda^4 \over 16 \pi^2}$  where $\lambda^4$ represents the product of
Yukawa couplings entering into the graph of Figure 12.

\end{itemize}

\begin{center}
\underline{\bf $W_{Higgs}$}
\end{center}

$$ +   {\bar \Psi}' A_2 \Psi  +  {\bar \Psi} A_2 \Psi' $$

$$ +   10_1  A_1 10_2  +  {\cal S}_7 10_2^2  $$

$$ +   {\bar \Psi} {\bar \Psi}' 10_1  +  \Psi \Psi' 10_1   $$

Note that the first two terms already appeared in the discussion of the
symmetry breaking sector.  They are included again here since as you
will see they are important for the discussion of the Higgs sector as
well.
It is the latter mechanism which has been entertained in the above
model.  As a result of these couplings to $\overline{\Psi}, \Psi$ the
Higgs doublets in $10_1$  mix with other states.  The mass matrix for
the SU(5) ${\bf \overline{5}, 5}$ states in ${\bf 10_1, 10_2,\Psi,
\Psi', \overline{\Psi}, \overline{\Psi}'}$ is given below.

\newpage

 \begin{eqnarray}  &\overline{5}_1   ~\overline{5}_2
{}~\overline{5}_{\Psi}  ~\overline{5}_{\Psi'} & \nonumber \\
\begin{array}{c} 5_1 \\ 5_2 \\ 5_{\overline{\Psi}} \\
5_{\overline{\Psi}'}
 \nonumber \end{array} &
\left( \begin{array}{cccc}  0 & A_1 & 0 & \Psi \\
                            A_1 & {\cal S}_7 & 0 & 0 \\
                            0 &  0 & 0 & A_2 \\
                            \overline{\Psi} & 0 & A_2 & 0 \end{array}
\right)&
\end{eqnarray}

{\bf Higgs doublets} In the doublet sector the vev $A_1$ vanishes.
Since the Higgs doublets in 10$_1$ now mix with other states, the
boundary condition $\lambda_b/\lambda_t = 1$ is corrected at tree
level.  The ratio is now given in terms of a ratio of mixing angles.

{\bf Proton decay} The rate for proton decay in this model is set by
the quantity  $(M^t)^{-1}_{11}$ where $M^t$ is the color triplet
Higgsino mass matrix.  We find  $(M^t)^{-1}_{11}= {{\cal S}_7 \over
A_1^2}$.  This may be much smaller than ${1 \over M_G}$ for ${\cal
S}_7$ sufficiently smaller than $M_G$.  Note there are no light color
triplet states in this limit.  Proton decay is suppressed since in this
limit the color triplet Higgsinos in $10_1$ become Dirac fermions (with
mass of order $M_G$), {\em but they do not mix with each other}.

\subsection{Symmetries} The theory has been constructed in order to
have enough symmetry to restrict the allowed operators.  This is
necessary in order to reproduce the mass operators in the effective
theory, as well as to preserve the vacuum directions assumed for the
45s and have natural doublet-triplet splitting.  Indeed the
construction of the symmetry breaking sector with the primed fields
allows the 45s to carry nontrivial U(1) charges.   This model has
several unbroken U(1) symmetries which do not seem to allow any new
mass operators.  It has a discrete $Z_4$ R parity in which all the
primed fields, ${\cal S}_{6,7}$ and $10_2$ are odd and $16_i, i =
1,2,3$ and $\overline{\Psi}_a, \Psi_a, a = 1, \cdots , 9$ go into $i$
times themselves. This guarantees that the odd states (and in
particular, $10_2$) do not couple into the fermion mass sector.  There
is in addition a Family Reflection Symmetry (see Dimopoulos- Georgi,
\cite{drw1}) which guarantees that the lightest supersymmetric particle
is stable.  Finally, there may be a continuous R symmetry which would
be useful in controlling Planck slop.  These conclusions are still
preliminary and must be checked.

{\bf Neutrino sector}  The neutrino sector seems to be very model
dependent.  It will constrain the symmetries of the theory, but I will
not discuss it further here.

\section{Conclusion}

In this talk, I have presented a class of supersymmetric SO(10) GUTs
which are in {\em quantitative} agreement with the low energy data.
With improved data these particular models may eventually be ruled out.
Nevertheless the approach of using low energy data to ascertain the
dominant operator contributions at $M_G$ seems robust.  Taking it
seriously, with {\em quantitative} fits to the data, may eventually
lead us to the correct theory.

Let me end this long talk with a short story.  I have been reading a
very interesting biography of Rutherford. It describes how, as a fresh
post-doc, Niels Bohr went to work with J.J. Thomson at the Cavendish
Laboratory.  He was interested in working with J.J. on a theory of the
atom. Bohr however didn't stay at the Cavendish for long; within the
year he left to go to Manchester to work with Rutherford.
Rutherford, of course, had recently proposed his theory of the atom to
explain how alpha particles could be scattered in the backward
direction upon hitting a thin metal foil.

The following excerpt, quoted from \underline{Rutherford : Simple
Genius} by D. Wilson, describes Bohr's feelings at the time.

In interviews later in his life Bohr explained that to Thomson, models
were mere analogies and fundamental problems of little interest.  {\it
Thomson did not demand consistency among the different models he
employed nor did he worry about quantitative agreement between
experiments and calculations based on the models}; indeed, in Bohr's
words `` things needed not to be very correct, and if it resembled a
little, then it was so"  as far as J.J. was concerned.

 The italics are mine.  I hope the moral of this story is evident.

\end{document}